\def\EuRb{EuRbFe$_4$As$_4$}
\def\cm{cm$^{-1}$}
\documentclass[aps,prl,lengthcheck,superscriptaddress,showpacs]{revtex4-1}
\usepackage{graphicx}%
\usepackage{color}
\usepackage{epstopdf}
\usepackage{amssymb}
\usepackage{amsmath}
\usepackage{amsfonts}

\begin{document}
\title{Unique interplay between superconducting and ferromagnetic orders in EuRbFe$_4$As$_4$}

\author{V. S. Stolyarov}
\email{vasiliy@travel.ru}
\affiliation{Moscow Institute of Physics and Technology, Dolgoprudny, 141700 Moscow, Russia}
\affiliation{Institute of Solid State Physics (ISSP RAS), 142432, Chernogolovka, Russia}
\affiliation{Kazan Federal University, 18 Kremlyovskaya str., Kazan 420008, Russia}

\author{A. Casano}
\affiliation{1. Physikalisches Institut, Universit{\"a}t Stuttgart, 70569 Stuttgart, Germany}

\author{M. A. Belyanchikov}
\affiliation{Moscow Institute of Physics and Technology, Dolgoprudny, 141700 Moscow, Russia}

\author{A. S. Astrakhantseva}
\affiliation{Moscow Institute of Physics and Technology, Dolgoprudny, 141700 Moscow, Russia}

\author{S. Yu. Grebenchuk},
\affiliation{Moscow Institute of Physics and Technology, Dolgoprudny, 141700 Moscow, Russia}

\author{D. S. Baranov}
\affiliation{Moscow Institute of Physics and Technology, Dolgoprudny, 141700 Moscow, Russia}
\affiliation{LPEM, ESPCI Paris, PSL Research University, CNRS, 75005 Paris, France}

\author{I. A. Golovchanskiy}
\affiliation{Moscow Institute of Physics and Technology, Dolgoprudny, 141700 Moscow, Russia}

\author{I. Voloshenko}
\affiliation{1. Physikalisches Institut, Universit{\"a}t Stuttgart, 70569 Stuttgart, Germany}

\author{E. S. Zhukova }
\affiliation{Moscow Institute of Physics and Technology, Dolgoprudny, 141700 Moscow, Russia}

\author{B. P. Gorshunov}
\affiliation{Moscow Institute of Physics and Technology, Dolgoprudny, 141700 Moscow, Russia}

\author{A. V. Muratov}
\affiliation{Lebedev Physical Institute, Russian Academy of Sciences, 119991 Moscow, Russia}

\author{V. V. Dremov}
\affiliation{Moscow Institute of Physics and Technology, Dolgoprudny, 141700 Moscow, Russia}

\author{L. Ya. Vinnikov}
\affiliation{Institute of Solid State Physics (ISSP RAS), 142432, Chernogolovka, Russia}

\author{D. Roditchev}
\affiliation{Moscow Institute of Physics and Technology, Dolgoprudny, 141700 Moscow, Russia}
\affiliation{LPEM, ESPCI Paris, PSL Research University, CNRS, 75005 Paris, France}
\affiliation{Sorbonne Universit\'e, CNRS, LPEM, 75005 Paris, France}

\author{Y. Liu}
\affiliation{Department of Physics, Zeijiang University, Hangzhou, 310027, China}

\author{G.-H. Cao}
\email{ghcao@zju.edu.cn}
\affiliation{Department of Physics, Zeijiang University, Hangzhou, 310027, China}

\author{M. Dressel}
\affiliation{Moscow Institute of Physics and Technology, Dolgoprudny, 141700 Moscow, Russia}
\affiliation{1. Physikalisches Institut, Universit{\"a}t Stuttgart, 70569 Stuttgart, Germany}

\author{E. Uykur}
\email{ece.uykur@pi1.physik.uni-stuttgart.de}
\affiliation{1. Physikalisches Institut, Universit{\"a}t Stuttgart, 70569 Stuttgart, Germany}

\date{\today}

\begin{abstract}

Transport, magnetic and optical investigations on EuRbFe$_4$As$_4$ single crystals evidence that the ferromagnetic ordering of the Eu$^{2+}$ magnetic moments at $T_N=15$~K, below the superconducting transition  ($T_c=36$~K), affects superconductivity in a weak but intriguing way. Upon cooling below $T_N$, the zero resistance state is preserved and the superconductivity is affected by the in-plane ferromagnetism mainly at domain boundaries; a perfect diamagnetism is recovered at low temperatures. The infrared conductivity is strongly suppressed in the far-infrared region below $T_c$, associated with the opening of a complete superconducting gap at $2\Delta = 10$~meV. A gap smaller than the weak coupling limit suggests the strong orbital effects or, within a multiband superconductivity scenario, the existence of a larger yet unrevealed gap.
 
\end{abstract}

\maketitle

New members of the iron-pnictide family, the so-called 1144-compounds, attract interest recently because
the alternating layers of alkaline $A$ and alkaline-earth $B$ cations produce two different kinds of As sites \citep{Hao2013, Kawashima2016, Iyo2016, Liu2016a}.
These materials can be viewed as the intergrowth of $A$-122 and $B$-122 iron-pnictides and they are naturally hole doped. The parent compounds are superconducting with transition temperatures $T_c$ around 35~K, higher than most of the 122-materials; no spin-density-wave order has been observed.
Among all possible candidates, Eu-based 1144-systems are even more intriguing, since the Eu-sublattice
orders ferromagnetically below a critical temperature $T_N\approx 15$~K \cite{Liu2016, Liu2017}, similar to the 122-counterpart EuFe$_2$As$_2$ \cite{Wu2009, Zapf2011, Zapf2013, Zapf2017, Maiwald2017}.
Ferromagnetic order deep inside the superconducting state is very rare, in general \cite{Sonin1998, Lorenz2005};
hence the ``ferromagnetic superconductor'' EuRbFe$_4$As$_4$ might pave the way towards realization
of a ``superconducting ferromagnet'' \cite{Nachtrab2006, Mineev2017, Jiao2017a}.
However, the exact nature of the Eu magnetic order and its effect on superconductivity
is unresolved \cite{Liu2016, Liu2017} because single crystals have been synthesized only recently.

In this Letter we focus on the interplay between superconductivity and ferromagnetism in \EuRb\ single crystals. We report comprehensive investigations comprising
transport, magnetic and optical measurements combined with microscopic studies of the vortex dynamics.
The infrared spectra show a clear gap opening around 80~\cm\ below $T_c=36$~K that is slightly reduced compared to the value expected from the BCS theory. We relate this small value to the multiband character of superconductivity as well as to the depairing (orbital) effects of supercurrents screening the ferromagnetic domains. A surprisingly weak effect on the superconducting condensate has been observed upon magnetic ordering indicating a rather weak interaction between Eu- and Fe- sublattices.

\begin{figure}
\centering
\includegraphics[width=1\linewidth]{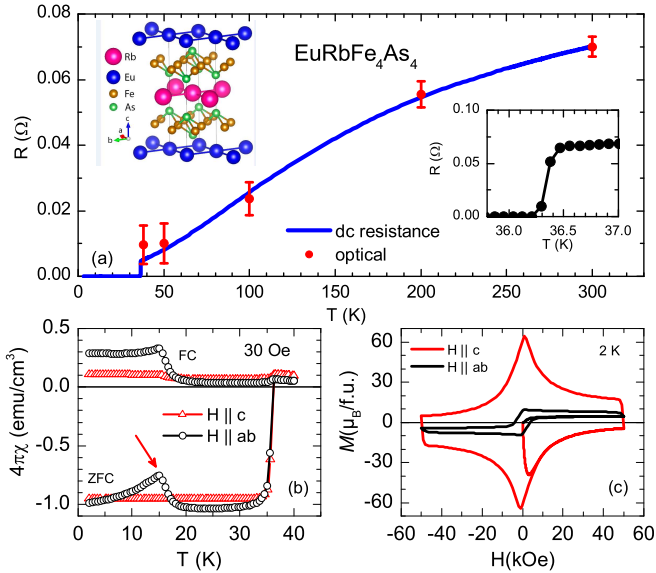}%
\caption{(a)~Temperature-dependent dc-resistance  (blue curve) of \EuRb\ single crystals measured within the $ab$-plane.
The red dots correspond to the resistivity extracted from the Hagen-Rubens fit of the reflectivity normalized to dc resistance 
at room temperature, with $\rho = 0.23~{\rm m}\Omega{\rm cm}$.
The sharp superconducting transition at $T_c=36.25$~K is magnified in the inset.
The structure of the unit cell illustrates the two spacing layers: Eu in blue and Rb in magenta. (b)~Magnetization at low temperatures measured by applying a magnetic field $H = 30$~Oe  within the $ab$-plane and perpendicular to it, $H\parallel c$.
The phase transitions are indicated by a sharp drop at superconducting $T_c$ and a pronounced feature around the magnetic order $T_N$, marked by a red arrow. The distinct behavior of field cooled (FC) and zero-field cooled (ZFC) susceptibility characterizes the superconducting state.
(c) The completely different field-dependent magnetization for both orientations hallmarks the confinement of the magnetic moments to the $ab$-plane.}%
\label{transport}%
\end{figure}

Single crystals of \EuRb\  are obtained according to Ref.~\cite{Liu2016,Liu2016a,Meier2017,Liu2018}; they exhibit shiny $ab$-faces of approximately 1~mm in size. The structure of the compound is presented in Fig.~\ref{transport}(a). The crystals are
characterized by x-ray, electrical transport, and magnetic susceptibility measurements.
In Fig.~\ref{transport}(a) we plot the dc resistance {\it vs.} temperature together
with the $\omega \rightarrow 0$ extrapolation of the infrared (IR) measurements.
The residual resistivity ratio (RRR) of 15.7 is significantly higher than reported for polycrystals \cite{Liu2016,Liu2017}.
The transition to the superconducting state takes place within a fraction of a degree [see inset in Fig.~\ref{transport}(a)]. No effect of the ferromagnetic ordering at $T_N=15$~K on resistivity is observed.

Fig.~\ref{transport}(b) displays the temperature dependence of the magnetization
probed in the field cooling (FC) and zero-field cooling (ZFC) protocol, for $H\parallel ab$ (black circles) and $H\parallel c$ (red triangles). The high quality of the crystals and the three-dimensional nature of the superconducting state are confirmed by the perfect diamagnetism rapidly reached below $T_c$ for both field orientations. Below $T_N$ the ferromagnetic ordering of the Eu-sublattice gives a kink-like anomaly in magnetic susceptibility (indicated by the red arrow) that is only visible for $H\parallel ab$. The absence of ferromagnetic signature for $H\parallel c$ is taken as evidence that the ferromagnetic order of the Eu-sublattice at $T_N = 15$~K results in domains oriented within the $ab$-plane.
The magnetization data presented in Fig.~\ref{transport}(c) confirm this conclusion: a pronounced hysteresis is observed within the plane, while the $c$-axis response is solely determined by the vortex dynamics \cite{Liu2018}. The behavior is very much in contrast to the phenomena observed in the EuFe$_2$As$_2$-family where canted A-type antiferromagnetism dominates with a reentrant spin glass behavior at lower temperatures \cite{Zapf2011,Zapf2013,Zapf2017}.

The $ab$-plane optical reflection measurements are performed in a frequency range
from 25 to 20\,000~\cm\ and down to $T=4$~K, using several Fourier-transform spectrometers complemented by IR microscopes and helium cryostats. The optical conductivity is obtained via the Kramers-Kronig analysis,
using  a Hagen-Rubens behavior in the normal state and Mattis-Bardeen fit in the superconducting state as low-frequency extrapolations; ellipsometric spectra collected up to 45\,000~\cm\ and x-ray scattering functions were supplemented at higher frequencies.
\begin{figure}
\centering
\includegraphics[width=1\columnwidth]{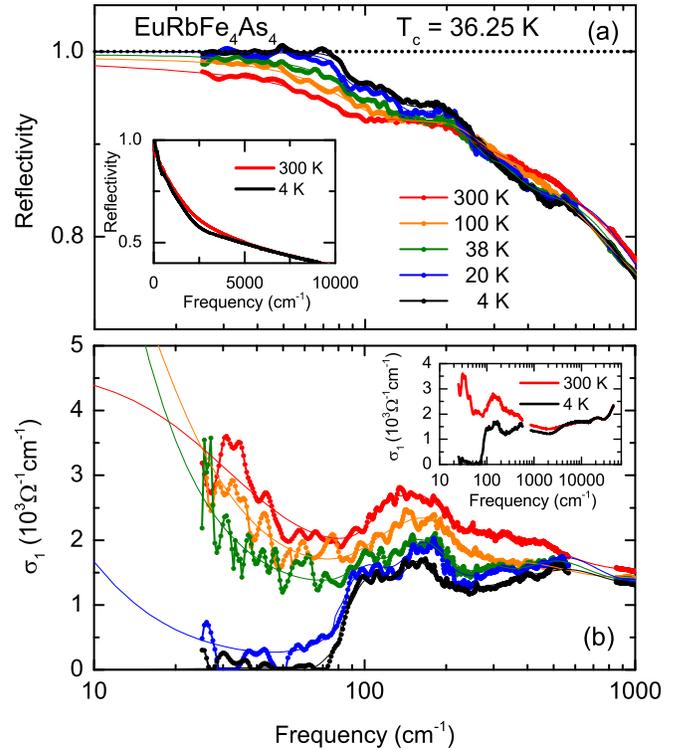}%
\caption{(a)~Temperature evolution of the infrared reflectivity of \EuRb; at the lowest temperature the spectra approach unity around 80~\cm. The inset displays the overall behavior at ambient and low temperatures. (b)~The corresponding conductivity spectra reveal the opening of the superconducting gap
when the temperature drops below $T_c=36$~K. The solid lines correspond to fits by the Drude-Lorentz model and Mattis-Bardeen equations, respectively. The inset shows the complete behavior at $T=300$ and 4~K.}%
\label{RefOC}%
\end{figure}
In Fig.~\ref{RefOC} the IR reflectivity and conductivity are plotted for selected temperatures. At a first glance the overall optical response of this new class of 1144-iron-pnictide resembles that of the 122-type systems \cite{Wu2010,Barisic2010,Nakajima2010,Uykur2015,Uykur2017}.

\begin{figure}
\centering
\includegraphics[width=1\columnwidth]{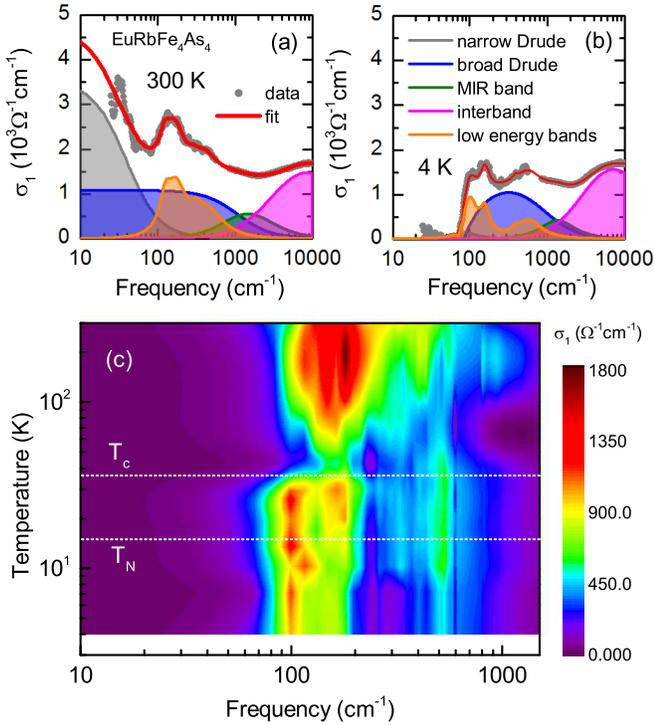}%
\caption{(a) Drude-Lorentz fits to the room-temperature optical conductivity of \EuRb. (b) In the superconducting state the two Drude modes are replaced by BCS terms \cite{DresselGruner02}. Note the drastic changes in the low-energy bands. (c) The temperature and frequency dependences of the low-lying absorption modes highlight the effects of the phase transitions.}%
\label{components}%
\end{figure}

In the normal state, optical data presented in Fig.~\ref{RefOC} can be described by two Drude components and series of Lorentzians, as demonstrated in Fig.~\ref{components}(a). The decomposition of the itinerant carriers into a narrow and a broad Drude term accounts for the multiband scenario of iron-pnictides \cite{Wu2010,Barisic2010}.
Similar to the observations in Eu-based 122-compounds \cite{Wu2009,Neubauer2016,Baumgartner2017}, the mid-IR band and higher-energy interband transitions are also visible.
Fig.~\ref{RefOC}(b) clearly shows that by decreasing $T$ the mid-IR absorption is suppressed and spectral weight transferred to higher energies; we interpret this behavior as an indication of Hund's rule coupling \cite{Schafgans2012}.
The low-energy absorption features [orange bands in Fig.~\ref{components}(a) and (b)] observed in \EuRb\ are unprecedented and their origin is not completely clear at this point.
The temperature evolution of these bands summarized in  Fig.~\ref{components}(c) reveals the splitting below $T_c$ and the redistribution at $T_N$.

The superconducting state is characterized by a distinct upturn of the reflectivity \cite{DresselGruner02} reaching unity at around 80~\cm. As a consequence, the optical conductivity drops and disappears for $\omega/(2\pi c) < 80$~\cm\ with basically no in-gap absorption. This evidences the presence of a complete gap throughout the Fermi surface, which might only have a small anisotropy \cite{Schachinger2011}.
The opening of the superconducting gap in $\sigma_1(\omega)$ can be well described by the Mattis-Bardeen equation \cite{Wu2010,Gorshunov2010} adding two
contributions, corresponding to the Drude terms in the normal state; the result is summarized in the solid lines of Figs.~\ref{RefOC}(b) and \ref{components}(b).
From the fits at 4~K we determine a gap value of $2\Delta_{0} = (80 \pm 5)~{\rm cm}^{-1} = (10 \pm 0.6)~{\rm meV} = (3.17 \pm 0.2) k_B T_c$ that is about 10$\%$ below predictions by weak-coupling theory.
The obtained frequency is in line with the small gap values detected for other 122-iron pnictides with a similar $T_c$ \cite{Dressel2011,Inosov2011}. Since in \EuRb\ several electronic bands cross the Fermi level, one may expect other gap(s) to be present, having the ratio $2\Delta_0/k_B T_c > 3.53$ \cite{Suhl1959}, similar to what is reported in 122 iron-based superconductors \cite{Dressel2011}.
For CaKFe$_4$As$_4$ with $T_c=31$-36~K, for instance, indications of two gaps, at 1-4~ meV and at 6-9~meV have been reported \cite{Biswas2017}.
The narrow Drude component of \EuRb, however, suggests that here we fall in the clean limit since $1/\tau_{nD} \approx 0.15 \Delta$; while the broad Drude term allows a small gap: $1/\tau_{bD} = 22 \Delta$ \cite{Dai2016, Neubauer2016}.

\begin{figure}
\centering
\includegraphics[width=0.9\columnwidth]{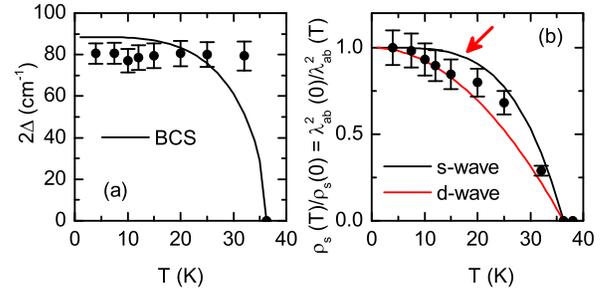}%
\caption{(a)~The superconducting gap of \EuRb\ deviates from the prediction of a mean-field  transition.
(b)~Temperature evolution of the superfluid density;
the $s$- and $d$-wave behavior is shown for comparison. A dip in the dependence is marked by red arrow}%
\label{lowlying}
\label{sd}
\end{figure}

It is remarkable that the gap $\Delta(T)$ in the IR spectra does not open gradually, as expected for a second-order transition, but rises sharply at $T_c$, as depicted in Fig.~\ref{sd}(a). Similar observations have been reported for other iron-pnictides such as electron-doped Eu122 \cite{Baumgartner2017}, Co-doped Ba122 \cite{Lobo2010} and Sm-1111 thin films \cite{Charnukha2018}. Further theoretical effort is required to
explain this behavior.

Upon the gap opening in the optical conductivity,
the missing area $A=\int[\sigma_1^n-\sigma_1^s]{\rm d}\omega$ between the normal and superconducting states, $\sigma_1^n$ and $\sigma_1^s$, respectively, is condensed to the $\delta$-function at $\omega = 0$,
according to the Ferrell-Glover-Tinkham sum rule \cite{DresselGruner02}.
Hence, the spectral weight analysis allows us to evaluate the penetration depth $\lambda = c/\sqrt{8A} \approx (420 \pm 45)$~nm  at $T=4$~K, comparable to similar iron-pnictides, and the superfluid density $\rho_s(T)\propto 1/\lambda^2(T)$; the results are consistent with calculations based on the imaginary component $\sigma_2(\omega)\propto \Delta_0/\hbar\omega$.
In Fig.~\ref{sd}(b) $\rho_s(T)$ is plotted together with predictions for isotropic ($s$-wave) and anisotropic ($d$-wave) gap symmetries.
The $s$-wave model seems more appropriate, consistent with the observed full gap opening in the IR spectra.

At first glance, there is no effect of the ferromagnetic order on superconductivity of \EuRb. Indeed, unlike in Eu-122 systems, where reentrant superconductivity has been reported \cite{Miclea2009,Paramanik2013,Jiao2017,Baumgartner2017}, 
we cannot detect any indication of ordering at $T_N$ on 
the resistivity curve [Fig.~\ref{transport}(a)] or any noticeable in-gap absorption due to vortices in IR spectra; instead, a fully developed superconducting gap and $s$-wave characteristic of the superfluid density is observed. However, several signatures of the ferromagnetic order are unveiled by our investigations. First, the magnetization curve in Fig.~\ref{transport}(b) measured with a magnetic field within the $ab$-plane shows a clear kink around $T=15$~K -- the perfect diamagnetism is destroyed; it is recovered only at significantly lower temperatures. Second, the superfluid density $\rho_s(T)$ in Fig.~\ref{sd}(b) manifests a weak dip in the range $T=10$-15~K, followed by the recovery of superfluid density at lower temperatures (red arrow). Both experimental results suggest that the superconductivity is substantially weakened at the ferromagnetic transition but somehow ``recovers'' at lower temperatures.

\begin{figure}
\centering
\includegraphics[width=1\linewidth]{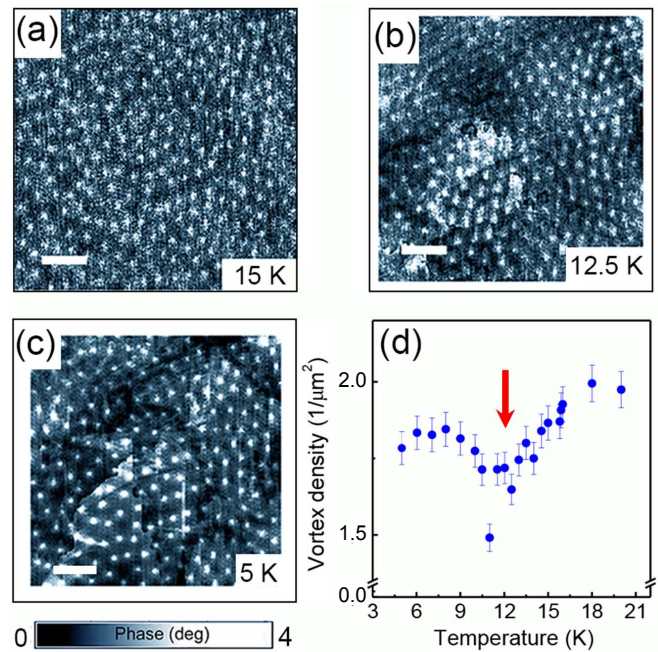}%
\caption{Abrikosov vortex lattice in \EuRb\ revealed by magnetic force microscopy. A field of $H=35$~Oe is applied along the $c$-direction, perpendicularly to the scanned sample surface.
(a) For $T_N < T < T_c$ the main magnetic contrast comes from quite uniformly distributed vortices. (b, c) For $T<T_N$ a large scale magnetic contrast due to ferromagnetic domain walls adds to the vortex signal; the positions of the vortices are modified. The white scale bars   correspond to 2~$\mu$m. (d) The temperature dependence of the vortex density integrated over the scanned area shows a minimum below the ferromagnetic transition, marked by red arrow. }
\label{vortices}%
\end{figure}

This ``weakening/recovery'' of superconductivity in \EuRb\ is confirmed by local MFM investigations, in which the influence of ferromagnetic ordering on the Abrikosov vortex lattice is studied. Since the MFM probes the $c$-component of the local magnetic field, the  vortex lattice is created by applying an external field along the $c$-axis of the crystal. The magnetic map in Fig.~\ref{vortices}(a), acquired just above $T_N$, shows a lattice of Abrikosov vortices emerging out of the $ab$-plane. The vortex organization, their surface density, and inter-vortex distances are typical of disordered type-II superconductors. The images displayed in Fig.~\ref{vortices}(b) and (c) correspond to the same region of the sample; the maps were acquired at $T=12.5$~K, right below $T_N$, and at $T=5$~K, respectively. In these maps an additional magnetic contrast appears on a larger spatial scale, which is associated with a $c$-oriented component of the local field due the domain walls between neighboring $ab$-plane oriented ferromagnetic domains. The fact that the vortices continue to exist below $T_N$ without any significant deformation of the vortex lattice evidences a rather weak local field generated by the ferromagnetic domains. This behavior is in contrast to the stark effect of ferromagnetism on superconductivity observed in other ferromagnetic superconductor EuFe$_2$(As$_{1-x}$P$_x$)$_2$ in which a $c$-oriented ferromagnetic ordering takes place, leading to novel magnetic superconducting phases \cite{Nowik2011,Veshchunov2017, Stolyarov2018}.

A detailed analysis of MFM data demonstrates that the ferromagnetic order does influences the vortex lattice, albeit weakly. Near $T_N$ the integrated vortex density is reduced as demonstrated in Fig.~\ref{vortices}(d). On a local scale, the vortex distribution becomes inhomogeneous: domains with unchanged vortex density coexist with regions where the density becomes significantly higher or lower. This indicates that below $T_N$ the vortex lattice is affected by ferromagnetism, but the intensity of the additional field remains rather low, as it is not able to alter the vortex lattice significantly. The weak coupling of ferromagnetism and superconductivity may imply the existence of a rather weak exchange fields between the Eu- and Fe- sublattices, where such effects have been discussed previously for the Eu-122 systems with Eu-bands located far away from the Fermi energy \cite{Jeevan2008}.  

From our investigations on single crystals, we conclude that \EuRb\ is a ferromagnetic multiband superconductor, in which superconductivity overcomes the $ab$-plane oriented ferromagnetic order of the Eu$^{2+}$ ions. A single excitation gap at $80~{\rm cm}^{-1}$ is observed below a sharp superconducting transition at $T_c = 36.25$~K; it reveals an unconventional temperature dependence. The gap energy $2\Delta_{0} = 3.17 k_B T_c$ is below the weak coupling limit, suggesting the existence of another larger superconducting gap. The reduction of the gap energy can also be associated with the depairing effect of spontaneous Meissner currents screening the ferromagnetic domains in the $ab$-plane below $T_N$.

\begin{acknowledgments}
We acknowledge the fruitful discussions with F. H\"{u}tt, D. G\"{u}nther, and A.V. Pronin and the technical support by G. Untereiner, and R. Aganisyan. V.S., D.B. and D.R. acknowledge the partial supports by the French Nat. Agency for Res. via grant SUPERSTRIPES and by the Ministry of Educ. and Sci. of the Russian Federation via grant 14.Y26.31.0007. 
D.R. acknowledge the COST project ``Nanoscale coherent hybrid devices for SC quantum technologies'' -- Action CA16218. The Magnetic Force Microscopy studies were funded by the Russian Sci. Fou. (project no 18-72-10118). Part of the work was supported by RFBR project 14-02-00255 and the Nat. Nat. Sci. Fou. of China (No. 11474252). B.G, E.Z.  thank support of the RAS Program of Fund. Res. ``Fundamental Problems of High-Temperature Superconductivity" and Program ``5-top100". We also acknowledge funding of the MIPT grant for visiting professors and of the DFG via DR228/42-1. E.U. acknowledges the support by the ESF and by the Ministry of Sci. Res. and the Arts Baden-W\"{u}rttemberg.
\end{acknowledgments}

\bibliographystyle{apsrev4-1}
\bibliography{EuRbFeAs_references}

\end{document}